\documentclass[12pt, preprint]{aastex}

\usepackage{natbib}

\newcommand{\teff}{${T}_{\mathrm{eff}}$}
\newcommand{\logg}{log $g$}
\newcommand{\msun}{${\mathrm{M}}_{\odot}$}
\newcommand{\wdo}{WD\,J1916+3938}

\shorttitle{A ZZ Ceti in the Kepler Field}
\shortauthors{Hermes et al.}

\begin{document}

\title{Discovery of a ZZ Ceti in the {\em Kepler} Mission Field}

\author{J. J. Hermes\altaffilmark{1,2}, Fergal Mullally\altaffilmark{3}, R. H. {\O}stensen\altaffilmark{4}, Kurtis A. Williams\altaffilmark{5}, John Telting\altaffilmark{6}, John Southworth\altaffilmark{7}, S. Bloemen\altaffilmark{4}, Steve B. Howell\altaffilmark{3}, Mark Everett\altaffilmark{8} and D. E. Winget\altaffilmark{1,2}}

\altaffiltext{1}{Department of Astronomy, University of Texas at Austin, Austin, TX\,-\,78712, USA}
\altaffiltext{2}{McDonald Observatory, Fort Davis, TX\,-\,79734, USA}
\altaffiltext{3}{NASA Ames Research Center, Moffett Field, CA\,-\,94035, USA}
\altaffiltext{4}{Instituut voor Sterrenkunde, K.U. Leuven, Celestijnenlaan 200D, B-3001 Leuven, Belgium}
\altaffiltext{5}{Department of Physics and Astronomy, Texas A\&M University -Commerce, Commerce, TX\,-\,75428, USA}
\altaffiltext{6}{Nordic Optical Telescope, 38700 Santa Cruz de La Palma, Spain}
\altaffiltext{7}{Department of Physics, University of Warwick, Coventry CV4 7AL, UK}
\altaffiltext{8}{National Optical Astronomy Observatory, Tucson, AZ\,-\,85719, USA}

\email{jjhermes@astro.as.utexas.edu}

\begin{abstract}
We report the discovery of the first identified pulsating DA white dwarf in the field of the {\em Kepler} mission, WD\,J1916+3938 (\textit{Kepler} ID 4552982). This ZZ Ceti star was first identified through ground-based, time-series photometry, and follow-up spectroscopy confirm it is a hydrogen-atmosphere white dwarf with \teff \, = 11,129 $\pm$ 115 K and \logg \, = 8.34 $\pm$ 0.06, placing it within the empirical ZZ Ceti instability strip. The object shows up to 0.5\% amplitude variability at several periods between 800 -- 1450 s. Extended {\em Kepler} observations of \wdo \, could yield the best lightcurve, to-date, of any pulsating white dwarf, allowing us to directly study the interior of an evolved object representative of the fate of the majority of stars in our Galaxy.
\end{abstract}


\section{Introduction}

Hydrogen atmosphere (DA) white dwarf stars account for more than 80\% of all spectroscopically identified WDs \citep{Eisenstein06}. When these objects cool to around 12,500 K, the hydrogen in their non-degenerate atmosphere forms a partial ionization zone, impeding energy transport and leading to global $g$-mode oscillations. It is believed that all DA white dwarf stars with carbon/oxygen cores undergo such pulsations when they reach the appropriate temperature \citep{Cast07}, allowing us to probe a stellar population representative of the fate of most stars in our Galaxy, including our Sun.

Asteroseismology provides a unique opportunity to investigate the internal structure of these stars (see reviews by \citealt{WinKep08,FontBrass08}). Given enough observed modes, we can put accurate estimates on the overall stellar mass, hydrogen and helium layer masses, internal chemical transition zones, magnetic field strength, and rotation rate. Performing asteroseismology on a WD benefits from its degenerate interior, which makes its structure less complicated than for stars still undergoing fusion.

In an effort to reduce aliasing caused by gaps in the data, much effort has been expended to obtain uninterrupted time-series observations of these variable DA (DAV) stars (also known as ZZ Ceti stars), using the Whole Earth Telescope \citep{Nather90}. However, these ground-based campaigns rarely extend beyond a few weeks, fundamentally limited by weather and moonlight, not to mention the availability of telescope time.

Space-based photometry has revolutionized the extended coverage of stars. The {\em Kepler} spacecraft was launched in March 2009, and while its primary mission is to detect Earth-sized planets around Sun-like stars, its extensive lightcurves are transforming many fields in asteroseismology. The spacecraft stares at some 156,000 stars and has the capability to monitor 512 objects with one-minute cadence. Month-long observations typically yield a duty cycle of more than 97\% with micromagnitude precision, depending on the brightness of the target.

Despite an expansive search \citep{Ostensen10,Ostensen11a} in which 17 compact objects were targeted by {\em Kepler} as potential pulsating white dwarf stars, none were observed to vary above the 4$\sigma$ detection limit. One DAV candidate, KIC10420021, was monitored for 7 months and is stable to at least 60 parts-per-million (ppm). \textit{Kepler} has monitored a uniquely variable DA, but that system is best explained by a magnetic spin-modulated WD model, not by pulsations \citep{HH11}. However, as part of a recent auxiliary search, \citet{Ostensen11b} announced the successful detection of the first pulsating WD in the \textit{Kepler} field, a V777 Her star. It was quite a fortuitous discovery, as this was just the second DB found in the field and has a \textit{Kepler} magnitude of $\textit{Kp}=18.46$.

Using facilities at the McDonald Observatory in west Texas, we have discovered the first ZZ Ceti star in the \textit{Kepler} field, WD J191643.83+393849.7 (\textit{Kepler} ID 4552982, hereafter \wdo). This object has a \textit{Kepler} magnitude of $\textit{Kp}=17.85$ \citep{Brown11}. Previous \textit{Kepler} observations of such faint, blue objects have reached better than 130 ppm (0.013\%) precision with just one month's worth of data. Extended short-cadence \textit{Kepler} observations will provide a detailed look at the pulsation spectrum of this cool ZZ Ceti star.


\section{High-Speed Photometric Observations}

\subsection{Selection and Time-Series Photometry}

Potential pulsating WDs in the \textit{Kepler} field were selected from a recent list of high reduced proper motion stars in the SuperCOSMOS Sky Survey \citep{Rowell11}. From their catalog of roughly 10,000 candidate WDs, we found 20 that actually fell on {\em Kepler} CCDs. Five of those had already been surveyed by \textit{Kepler} and found not to vary \citep{Ostensen10,Ostensen11a}.

One of the remaining 15, \wdo, fell in the footprint of the POSS-II survey, and showed ($\textit{B}-$\textit{R}), ($\textit{R}-$\textit{I}) colors extremely close to values for a typical ZZ Ceti star, so it was targeted for time-series photometric observations. The discovery observations were taken at the McDonald Observatory using the Argos instrument, a frame-transfer CCD mounted at the prime focus of the 2.1m Otto Struve telescope \citep{Nather04}. Data were obtained through a 1mm BG40 filter to reduce sky noise.


Photometric observations were carried out over 6 nights in May 2011, for a total of more than 21 hours of coverage. Exposure times ranged from 5 to 15 seconds, depending on seeing and cloud conditions.

We performed weighted, circular aperture photometry on the calibrated frames using the external IRAF package $\textit{ccd\_hsp}$ written by Antonio Kanaan. We divided the sky-subtracted lightcurves using at least 3 brighter comparison stars in the field to allow for fluctuations in seeing and cloud cover. Using the WQED software suite \citep{Thompson09}, we fit a second-order polynomial to the data to remove the long-term trend caused by atmospheric extinction, and applied a timing correction to each observation to account for the motion of the Earth around the barycenter of the solar system.

\begin{figure}[t]
\centering{\includegraphics[width=\columnwidth]{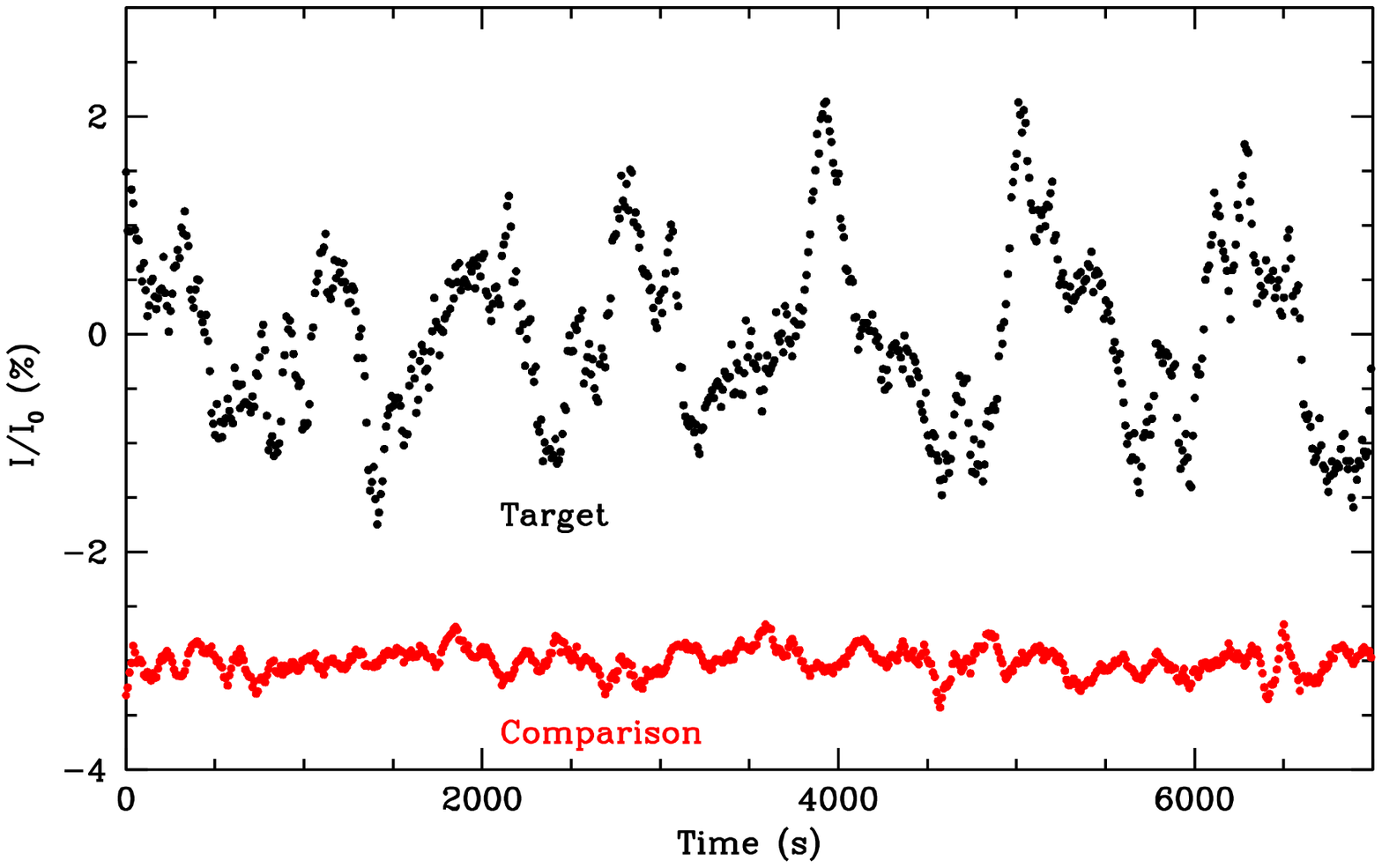}}
\caption{A portion of the lightcurve for WD\,J1916+3938, taken on 2011 May 11, that has been smoothed by a four-point moving average. The red, offset lightcurve is of the brightest comparison star in the field. \label{fig1}}
\end{figure}

Figure~\ref{fig1} shows a representative lightcurve, smoothed by a four-point moving average window for visualization purposes with the brightest comparison star offset below.


\subsection{Ground-Based Lightcurve Analysis}

\begin{figure*}[t!]
\plotone{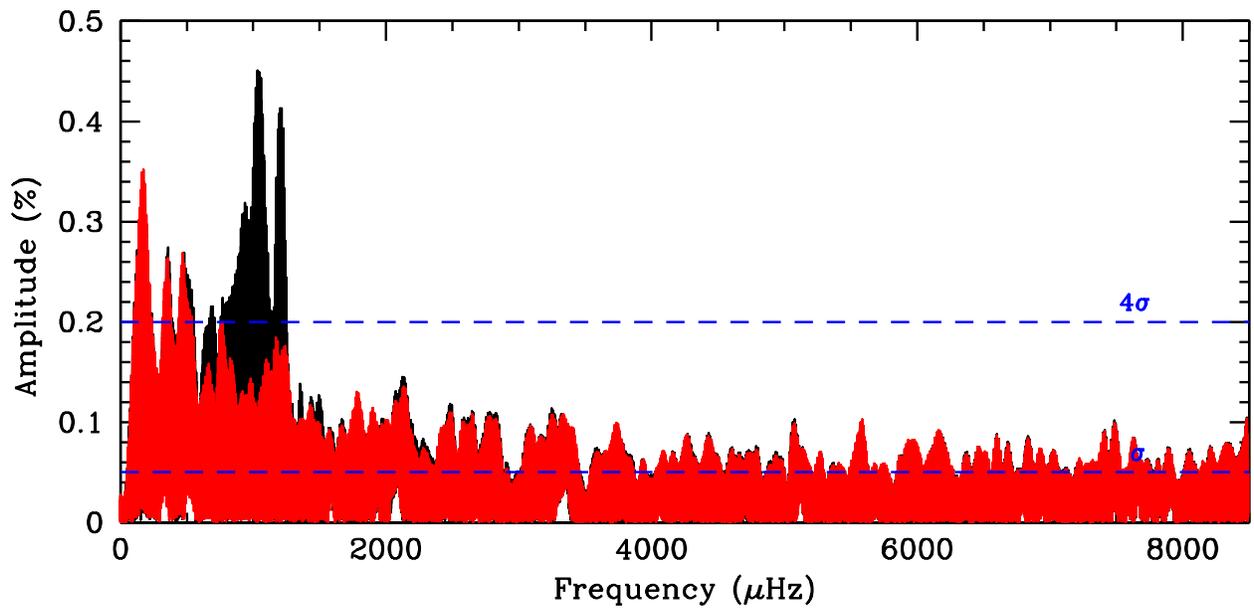}
\caption{A Fourier transform of our May 2011 data on WD\,J1916+3938. The red transform has been computed after prewhitening by the seven periodicities listed in Table~\ref{freq}, which removes all peaks above four times the mean FT level, $\sigma$. We have ignored the low-frequency peaks in this FT, with periods longer than 2,000 s, which are likely to be noise from variability in the Earth's atmosphere. \label{fig2}}
\end{figure*}

We performed a Fourier transform of the entire data set, 7,454 points spread over nearly a month for a duty cycle of 3.1\% (Figure~\ref{fig2}). This FT extends to the Nyquist frequency for \textit{Kepler} short-cadence observations; there is nothing above the 4$\sigma$ threshold for higher frequencies. The window function is quite messy, as there are variably spaced gaps in the data.


While \textit{Kepler} will allow us to improve our determination of the periodicities present in this star, we attempt to identify the periods present in all our May 2011 data. Table~\ref{freq} lists the highest seven peaks, all with power above four times the mean FT level. The large aliasing present makes it nearly impossible to determine, with much certainty, the true periods in this short, ground-based data set, so we have omitted the formal least-squares errors. It appears this star has several excited modes between 800 -- 1450 s, although the amplitudes appear unstable from night to night. There is consistently power well above the noise around 970 s and 820 s each night, although the level of excitation appears inconsistent.


For comparison sake, we also took an FT of all the data for the brightest comparison star in the field. The average amplitude over all frequency space is 0.03\%, the same level as our formal least-squares amplitude errors.

\section{Spectroscopic Analysis}

\subsection{Spectroscopic Observations}

Excitement among the {\em Kepler} compact object community, specifically Working Group 11 within the Kepler Asteroseismic Science Consortium (KASC), led to multiple spectroscopic follow-ups, and \wdo \, was observed using 4 different telescopes, especially after our first spectrum from the Nordic Optical Telescope (NOT) confirmed the object as a DA within the ZZ Ceti instability strip. Fits to these observations were used to derive an effective temperature and surface gravity.

Observations using the 9.2m Hobby-Eberly Telescope (HET) at McDonald Observatory were carried out with the Marcario Low Resolution Spectrograph (LRS) using its g2\_2.0 setup. The observations using the 2.6m NOT at Roque de los Muchachos Observatory were done in low-resolution with the ALFOSC spectrograph and its grism \#7. The observations using the 3.5m telescope at Calar Alto Observatory (CAHA) were done in medium-resolution with the TWIN spectrograph, using grating T06 for the red arm and grating T12 for the blue arm. Finally, the observations using the 4m Mayall Telescope at Kitt Peak National Observatory (KPNO) were done in low-resolution with the R-C CCD Spectrograph. Each spectral image and flux standard were reduced using standard long-slit IRAF routines.

\begin{figure*}[t!]
\plotone{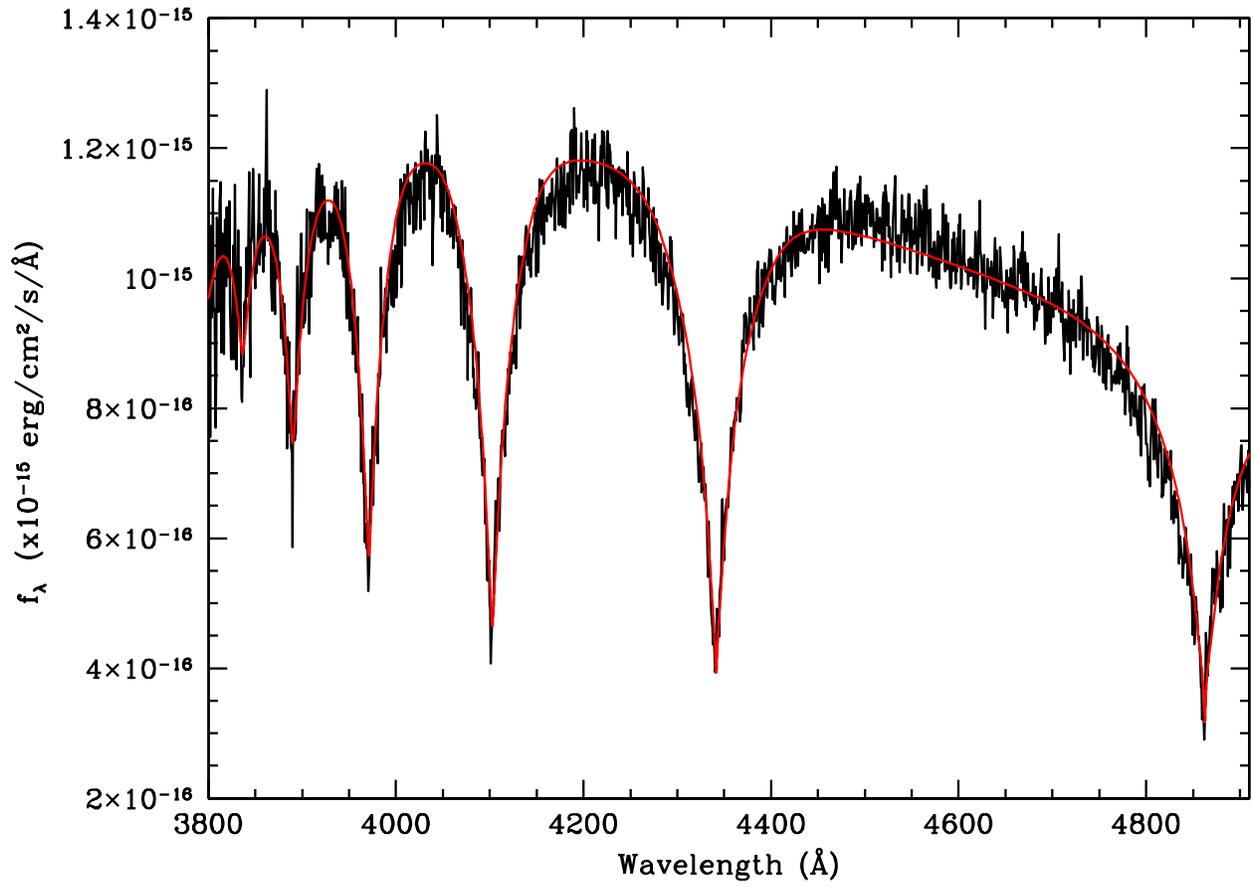}
\caption{The KPNO spectrum of WD\,J1916+3938, showing the Balmer lines H$_{\beta}$ through H$_{9}$. Overplotted in red is the best-fit DA model spectrum. \label{fig3}}
\end{figure*}

The KPNO spectrum of \wdo, showing the Balmer lines H$_{\beta}$ through H$_{9}$, can be seen in Figure~\ref{fig3}. Overplotted is the best-fit DA model spectrum.


We have fit model grid spectra to each of these observations, using the procedure outlined in \citet{Bergeron92}. The WD models used for the spectroscopic fitting were kindly provided by Detlev Koester and are described in \citet{Koester10}. Balmer lines in the models were calculated with the modified Stark broadening profiles of \citet{TB09}, kindly made available by the authors, and use the mixing length prescription ML2/$\mathrm{\alpha}$=0.6. Table~\ref{spec} lists the best fits to the observations. Using a weighted mean of all four determinations, we adopt a best value of \teff \, = 11,129 $\pm$ 115 K and \logg \, = 8.34 $\pm$ 0.06. This places the object within the empirical ZZ Ceti instability strip.

\subsection{Implications}

ZZ Ceti stars at the hotter, blue edge of the instability strip tend to pulsate with short period (100 -- 300 s) modes that have shown to be extremely stable in period and amplitude. Such stars have been the focus of work to detect an evolutionary rate of period change of these modes (e.g., \citealt{Kepler05,Mukadam09}). However, the ZZ Ceti stars near the cooler, red edge of the instability strip with effective temperatures near 11,000 K typically feature much longer periods ($>$ 600 s) and have more excited modes. These pulsations are typically unstable in amplitude and phase.

Spectroscopic fits show that \wdo \, has an effective temperature that places it in the cooler region of the ZZ Ceti instability strip. This result is duly confirmed by our time-series photometric observations. \wdo \, shows long-period oscillations --- its weighted mean period \citep{Mukadam06} lies around 1,003 s --- that do not appear stable in amplitude. For example, during the runs from May 27--29, the power observed in the mode near 824 s went from (0.66 $\pm$ 0.06)\% to (0.88 $\pm$ 0.05)\% to (0.43 $\pm$ 0.05)\%, respectively.

Many modes in other cool ZZ Ceti stars have amplitudes that grow and decay in a way that cannot be explained by beating (e.g., \citealt{Kleinman98}), so we expect the same is true with this star. The opportunity for nearly continuous {\em Kepler} observations of this amplitude modulation may afford us a completely new insight into the growth and decay of mode amplitudes, opening a new window into how these objects behave on a variety of timescales inaccessible from the ground.

\section{Discussion and Conclusion}

We have discovered just the second pulsating white dwarf in the {\em Kepler} field, and the first ZZ Ceti star suitable for {\em Kepler} monitoring. Several modes appear to be excited from 800 -- 1450 s, with a fractional amplitude of up to 0.5\%. Spectroscopic fits yield an effective temperature of 11,129 $\pm$ 115 K and a surface gravity of \logg \, = 8.34 $\pm$ 0.06, which corresponds to a mass of 0.82 $\pm$ 0.04 \msun \,from the evolutionary models of \citet{Wood90}. This spectroscopic temperature puts it on the cool, red edge of the ZZ Ceti instability strip, where pulsations may begin shutting down.

Submitted for {\em Kepler} observations through KASC, the spacecraft is scheduled to commence short-cadence observations on this ZZ Ceti pulsator in October 2011 (Q11). By virtue of its impressive duty cycle, we anticipate that extended {\em Kepler} observations will yield the best lightcurve, to date, of any pulsating white dwarf, even those that are more than four magnitudes brighter.

The lightcurves can be analyzed from a variety of asteroseismic angles. With enough modes, we will be able to put an accurate constraint on the overall mass of the WD. Perhaps more importantly, we may also extract an accurate estimate for the hydrogen layer mass, which controls the rate at which this star cools. This hydrogen mass fraction has been constrained through asteroseismology in fewer than a dozen DA WDs \citep{Dolez06}, although \citet{Cast09} attempted to derive the parameter from the few modes present in an ensemble of 83 ZZ Ceti stars. Uncertainties in this parameter is imprinted as scatter in the white dwarf luminosity function, which is used in cosmochronology to put an estimate on the age of the Galaxy (e.g., \citealt{Winget87, Harris06}).

The sensitivity of {\em Kepler} observations may also allow us to see a wealth of low-amplitude multiplet structure. Previous ground-based campaigns on cool ZZ Ceti stars, such as G29-38, have revealed instability in the amplitude of these multiplets \citep{Kleinman98}. Extended {\em Kepler} coverage will allow us to observe how the amplitudes of all modes in this star change with time.

ZZ Ceti stars often exhibit some nonlinear combination frequencies, which are believed to arise from convective driving (e.g., \citealt{Brickhill92, Wu01}). If observable, these combination frequencies may be used to identify the spherical degree of the parent modes \citep{Yeates05}. With an ideal {\em Kepler} lightcurve, we would be able to compare this mode identification with the $\ell$ values as determined from the period spacings. Pulse-shape fitting of properly identified modes may also be used to constrain some of the convective parameters in this star, especially the depth of the convection zone \citep{Mikemon10}.

For now, though, we eagerly await the unrivaled staring competition \textit{Kepler} will have with this cool ZZ Ceti star.

\acknowledgments
We thank KASC WG11 for their enthusiasm, especially regarding spectroscopic follow-ups. This work is supported by the Norman Hackerman Advanced Research Program, under grants 003658-0255-2007 and 003658-0252-2009, by a grant from the NASA Origins Program, NAG5-13094, and by the National Science Foundation, under grant AST-0909107. The research leading to these results has received funding from the European Research Council under the European Community's Seventh Framework Programme (FP7/2007--2013)/ERC grant agreement n$^\circ$227224 (PROSPERITY), as well as from the Research Council of K.U.Leuven grant agreement GOA/2008/04. This work was partly based on observations made with the Nordic Optical Telescope, operated on the island of La Palma jointly by Denmark, Finland, Iceland, Norway, and Sweden. We thank the McDonald Observatory staff for their support, especially Dave Doss and John Kuehne.

\clearpage

\clearpage

\begin{table}
  \caption{Frequencies present in WD\,J1916+3938 from May 3--31.}
  \label{freq}
  \begin{center}
    \leavevmode
    \begin{tabular}{cccc} \hline \hline              
ID & Freq. ($\mu$Hz) & Period (s) & Amp. (\%) \\ \hline 
$f_{1}$ & 1032.1 & 968.9 & 0.44 \\
$f_{2}$ & 1213.7 & 823.9 & 0.38 \\
$f_{3}$ & 1070.1 & 934.5 & 0.36 \\
$f_{4}$ & 1198.9 & 834.1 & 0.32 \\
$f_{5}$ & 918.3 & 1089.0 & 0.25 \\
$f_{6}$ & 696.0 & 1436.7 & 0.24 \\
$f_{7}$ & 854.8 & 1169.9 & 0.23 \\ \hline
    \end{tabular}
  \end{center}
\end{table}

\begin{table}
  \caption{Spectroscopically determined parameters.}
  \label{spec}
  \begin{center}
    \leavevmode
    \begin{tabular}{lcccc} \hline \hline              
Instrument & Date Observed & Wavelength Range (\AA) & \teff \,(K) & \logg \, (dex) \\ \hline 
NOT/ALFOSC & 2011-05-22 & 3850--6850 & 11,152 $\pm$ 80 & 8.24 $\pm$ 0.06 \\
CAHA/TWIN & 2011-06-17 & 3500--5450/5700--6790 & 11,116 $\pm$ 139 & 8.36 $\pm$ 0.09 \\
HET/LRS & 2011-06-14 & 4300--7200 & 11,104 $\pm$ 150 & 9.51 $\pm$ 0.43 \\
KPNO/RCS & 2011-06-09 & 3650--5120 & 11,115 $\pm$ 92 & 8.43 $\pm$ 0.05 \\
{\bf Adopted } & & & {\bf 11,129 $\pm$ 115} & {\bf 8.34 $\pm$ 0.06 } \\ \hline
    \end{tabular}
  \end{center}
\end{table}

\end{document}